\documentclass[12pt]{article}

\def\unnumbered{\setcounter{secnumdepth}{0}}%

\usepackage{arxiv}

\usepackage{hyperref}
\usepackage{floatpag}
\usepackage{graphicx}%
\usepackage{multirow}%
\usepackage{amsmath,amssymb,amsfonts}%
\usepackage{amsthm}%
\usepackage{mathrsfs}%
\usepackage[title]{appendix}%
\usepackage{xcolor}%
\usepackage{textcomp}%
\usepackage{manyfoot}%
\usepackage{booktabs}%
\usepackage{algorithm}%
\usepackage{algorithmicx}%
\usepackage{algpseudocode}%
\usepackage{listings}%
\usepackage{tabularray}
\usepackage{tabularx}
\usepackage{float}
\usepackage{graphicx}
\usepackage{caption}
\usepackage{subcaption}
\usepackage{lineno}
\usepackage{authblk}

\UseTblrLibrary{booktabs}
\UseTblrLibrary{siunitx}

\NewTableCommand{\tinytableDefineColor}[3]{\definecolor{#1}{#2}{#3}}

\unnumbered
\date{}

\renewcommand{\shorttitle}{Population-Scale Network Embeddings and Right-Wing Populist Voting}

\title{\nolinenumbers Population-Scale Network Embeddings Expose Educational Divides in Network Structure Related to Right-Wing Populist Voting}

\author[1,2,3\thanks{Corresponding author. Email: \href{mailto:m.luken@esciencecenter.nl}{m.luken@esciencecenter.nl}.}]{Malte Lüken}

\author[4,5]{Javier Garcia-Bernardo}

\author[6]{Sreeparna Deb}

\author[1,3]{Flavio Hafner}

\author[7]{Megha Khosla}

\affil[1]{Netherlands eScience Center, The Netherlands}

\affil[2]{Department of Psychology, University of Amsterdam, The Netherlands}

\affil[3]{Erasmus School of Behavioral and Social Sciences, Erasmus University Rotterdam, The Netherlands}

\affil[4]{ODISSEI Social Data Science Team, Department of Methodology and Statistics, Utrecht University, The Netherlands}

\affil[5]{Centre for Complex System Studies, Utrecht University, The Netherlands}

\affil[6]{Department of Software Technology, Delft University of Technology, The Netherlands}

\affil[7]{Department of Intelligent Systems, Delft University of Technology, The Netherlands}

\begin{document}

\maketitle

\begin{abstract}
    Administrative registry data can be used to construct population-scale networks whose ties reflect shared social contexts between persons. With machine learning, such networks can be encoded into numerical representations---embeddings---that automatically capture an individual's position within the network. We created embeddings for all persons in the Dutch population from a population-scale network that represents five shared contexts: neighborhood, work, family, household, and school. To assess the informativeness of these embeddings, we used them to predict right-wing populist voting. Embeddings alone predicted right-wing populist voting above chance-level but performed worse than individual characteristics. Combining the best subset of embeddings with individual characteristics only slightly improved predictions. After transforming the embeddings to make their dimensions more sparse and orthogonal, we found that one embedding dimension was strongly associated with the outcome. Mapping this dimension back to the population network revealed that differences in educational ties and attainment corresponded to distinct network structures associated with right-wing populist voting. Our study contributes methodologically by demonstrating how population-scale network embeddings can be made interpretable, and substantively by linking structural network differences in education to right-wing populist voting.
\end{abstract}


\section{Introduction}

Combining administrative registry data with machine learning has opened new opportunities for social science research \cite{hocuk_economies_2021, kunaschk_enriching_2024, narayanan_using_2025, kuikka_unpredictability_2024, hansen_predicting_2023}. A recent example is life2vec, a deep learning model trained on life sequences from Danish registry data spanning the entire Danish population that substantially improved predictions of life events such as early mortality \cite{savcisens_using_2023}. Another recent example is the PreFer data challenge, in which researchers used a combination of registry and survey data from the Dutch population to predict childbirth at the individual level \cite{sivak_combining_2024}.

To make accurate predictions, deep learning models typically encode input data into latent numerical representations---so-called embeddings \cite{ericsson_selfsupervised_2022,bengio_representation_2013}. Their goal is to reduce the dimensionality while also preserving the structure and non-linear relationships of complex input data. In contrast to hand-crafted features, embeddings capture the structure of complex inputs automatically from the data. Within the social sciences, this is especially relevant in networks \cite{zhang_network_2020} since the position of persons within social networks impacts social outcomes (e.g., economic \cite{granovetter_impact_2005}, health-related \cite{smith_social_2008}, and political \cite{lazer_networks_2011}), and capturing such position using hand-crafted features is particularly difficult \cite{stulp_datadriven_2023}.   

Registry data contain information about shared social contexts between persons that can be used to construct networks at population-scale \cite{vanderlaan_whole_2023, cremers_unveiling_2024, panayiotou_anatomy_2025}. For example, when people live in the same household or work for the same employer, this shared social context can be indicated by a network tie. While such ties do not measure direct interaction, they reflect an increased probability of contact and shared exposure to contextual factors \cite{bokanyi_anatomy_2023}. In this study, we draw on a recently released network based on Dutch registry data comprising \emph{all} persons living in the Netherlands  \cite{vanderlaan_whole_2023}. This network records relational ties representing five social contexts, namely neighbors, colleagues, family, classmates, and household, for 17.4 million residents, totaling 1.4 billion connections.

Using the population network, we created embeddings for the entire Dutch population solely based on shared social contexts (Fig.~\ref{fig:workflow}A). The embeddings encode each person's position within the network, capturing complex connectivity patterns and structural roles in a compact representation. The goal of this study was to gain insight into the information the embeddings contain about a relevant social outcome---right-wing populist voting (Fig.~\ref{fig:workflow}B)---and the structural patterns in the network associated with it (Fig.~\ref{fig:workflow}C). 

\subsection{Social Networks and Right-Wing Populist Voting}

The Dutch parliamentary election on November 22nd, 2023, yielded an unprecedented high turnout for right-wing populist parties \cite{vanholsteyn_dutch_2025}. For the first time in Dutch history, a right-wing populist party (\textit{Partij voor de Vrijheid}; PVV) was part of the government coalition.

Multiple theories explain the rise of (right-wing) populism through the lens of social networks. First, the mass society hypothesis states that individuals in modern societies have fewer meaningful social ties and are socially more isolated \cite{rydgren_legacy_2011}. This ``loss of community'' has in turn increased the potential for embracing populist ideologies that typically promote a homogeneous society. Recent lines of research in political psychology have also documented an association between decreased feelings of social belonging and right-wing populist voting (e.g., \cite{langenkamp_populism_2022}). One recent study found support for the mass society hypothesis in the Dutch population network, demonstrating that social cohesion---measured as average network closure---has declined by more than 15\% between the years 2010 and 2021 \cite{bokanyi_fragmentation_2026}.

A second theory, social capital theory \cite{bourdieu_forms_1986, portes_social_1998}, focuses on resources that are provided through social networks. This includes economic means, information, but also norms, obligations, and values. The exchange of these resources creates mutual trust and makes it easier for people to coordinate their actions \cite{putnam_making_1993}. Putnam \cite{putnam_bowling_2000} centers social capital around voluntary organizations (e.g., sports clubs) that promote the spread of civic values and institutional trust, and can thus be assumed to lower the appeal of populist ideologies. More recent studies on social capital have also highlighted the importance of informal social ties, such as friendship networks (e.g., \cite{stauder_friendship_2014, chetty_social_2022}). The role of social capital has also been investigated in the Dutch population network, showing that ties to affluent others and diverse social networks (``bridging'' capital) are important for predicting economic mobility, especially among disadvantaged social groups \cite{kazmina_can_2025}. 

A third theory, cleavage theory, states that voting behavior is shaped by societal divisions, conflicts, and structures that are aligned with party politics \cite{lipset_cleavage_1967, bornschier_cleavage_2024}. Traditionally, political cleavages were assumed to follow differences in class or religion. However, with the rise of populism, the link between traditional cleavage dimensions and voting behavior has been questioned \cite{bornschier_cleavage_2024}: A debate has emerged between \textit{de}-alignment of societal divisions, party politics and voting, and \textit{re}-alignment with new cleavage dimensions (e.g., education, urban-vs-rural). By drawing on social capital theory and theories of social closure, political cleavages have also been linked to social networks \cite{westheuser_cleavage_2025}. The assumption is that cleavage dimensions are aligned with network homogeneity, resulting in segregation. Socio-economic segregation has also been found in the Dutch population network, with the strongest assortativity along income, education, and urban-vs-rural dimensions \cite{vanderlaan_whole_2023, kazmina_socioeconomic_2024}. 

From the perspective of cleavage theory and network segregation, right-wing populist voters are more likely to have social ties with others who share similar ideology and socio-economic characteristics. Here, by investigating differences in network structure associated with right-wing populist voting, we can identify potentially novel cleavage dimensions or find support for dimensions that have already been proposed. This requires measuring (local) network structure and relating it to the outcome of interest.

\subsection{Measuring Social Network Structure}

Traditionally, social network structure has been measured with surveys (e.g., \cite{knoke_networks_1990, nieuwbeerta_crosscutting_2000, attewell_educational_2025, dejong_separated_2025}). While they are effective for measuring the most important informal social ties (e.g., close friendships or family relations), surveys typically only capture a small part of a person's ego network and miss connections among alters as well as weak, long-range ties. More recent studies have relied on mobile communication or social media networks that cover larger samples, however, these data suffer from coverage biases, non-representativity, and tend to lack contextual information \cite{bokanyi_fragmentation_2026}.

Population-scale administrative data provide a broad picture of formal social ties that reflect shared exposure to contextual factors and social opportunities \cite{bokanyi_anatomy_2023,vanderlaan_whole_2023}. The Dutch population network has been shown to have similar connectivity and community structure to a Dutch social media network with communities being centered at metropolitan areas or clustering along interdependent municipalities in the countryside \cite{menyhert_connectivity_2025}. For an in-depth explanation of the construction of the Dutch population network and its ``anatomy'' we refer to previous studies (\cite{bokanyi_anatomy_2023}; see also \cite{cremers_unveiling_2024, panayiotou_anatomy_2025} for descriptions of population-scale networks in Denmark and Sweden). Important for our study is that the Dutch population-scale network is extremely large and dense: collapsed across all layers, it contains 17M nodes and 1.4B edges, the median degree is 61, the diameter 22, and the shortest path length 4.64 \cite{bokanyi_anatomy_2023,vanderlaan_whole_2023}. This makes it challenging to effectively capture the (local) network structure, especially with handcrafted network statistics that have been used by previous studies to assess the importance of social ties for predicting various outcomes, including educational attainment \cite{garcia-bernardo_prediction_2025}, the spread of COVID-19 \cite{garcia-bernardo_assessing_2024,hedde-vonwesternhagen_predicting_2024}, and attitudes towards immigrants \cite{kazmina_contact_2024}. Similar to some of these studies, we linked the Dutch population network with representative survey data. This allowed us to assess the importance of local network structure for predicting right-wing populist voting behavior.

\subsection{Node Embeddings Through Sampling From Population-Scale Networks}

Instead of measuring network structure with handcrafted statistics, we represented persons in terms of their broader relational context by using node embeddings. Embedding methods capture structural similarities between persons that may not be visible from direct ties alone \cite{zhang_network_2020}. We employed two sampling-based network embedding methods, DeepWalk and LINE, because they are computationally efficient at population scale and have shown strong practical performance in downstream tasks \cite{khosla_comparative_2021, zhao_multilabel_2024}. Embedding methods can be understood as non-parametric latent space models because they map each individual (node) to a low-dimensional vector representation where distance reflects relational similarity \cite{ericsson_selfsupervised_2022, bengio_representation_2013}. A related way of obtaining low-dimensional vector representation of networks are parametric space models, such as the ideal point model. These models are typically used to estimate the ideology of politicians interacting with bills (e.g., \cite{poole_spatial_1985}) or attitudes of social media users interacting with each other (e.g., \cite{barbera_birds_2015, ramaciottimorales_inferring_2022}). Both network embeddings and parametric latent space models are
different computational strategies for the same underlying task: low-rank
approximation of matrices encoding network structure (both parametric latent space models and DeepWalk and LINE are equivalent to different types of matrix factorization \cite{barbera_birds_2015, qiu_network_2018}). The efficiency of
sampling-based methods comes from stochastic optimization over node pairs rather than optimization over the full graph. For example, DeepWalk \cite{perozzi_deepwalk_2014} generates short random walks from each node and treats these walks like sentences in language modeling. It then applies the Skip-gram model \cite{mikolov_distributed_2013}, where the goal is to predict nearby nodes from a given node, analogous to predicting nearby words from a word in a sentence. In this way, nodes that co-occur in short walks receive similar vector representations, allowing DeepWalk to capture both local connectivity and broader graph structure. The LINE (large-scale information network embedding; \cite{tang_line_2015}) method, in contrast, relies on edge-based sampling to efficiently preserve local proximity and neighborhood relationships. Using these sampling-based methods makes it possible to create node embeddings for a large and dense population-scale network. The quality of these embeddings may depend on hyperparameters such as walk length, context window size, negative sampling rate, and the number of embedding dimensions.

\subsection{Making Node Embeddings Interpretable}

\begin{figure}
    \centering
    \floatpagestyle{plain}
    \includegraphics[width=1\textwidth]{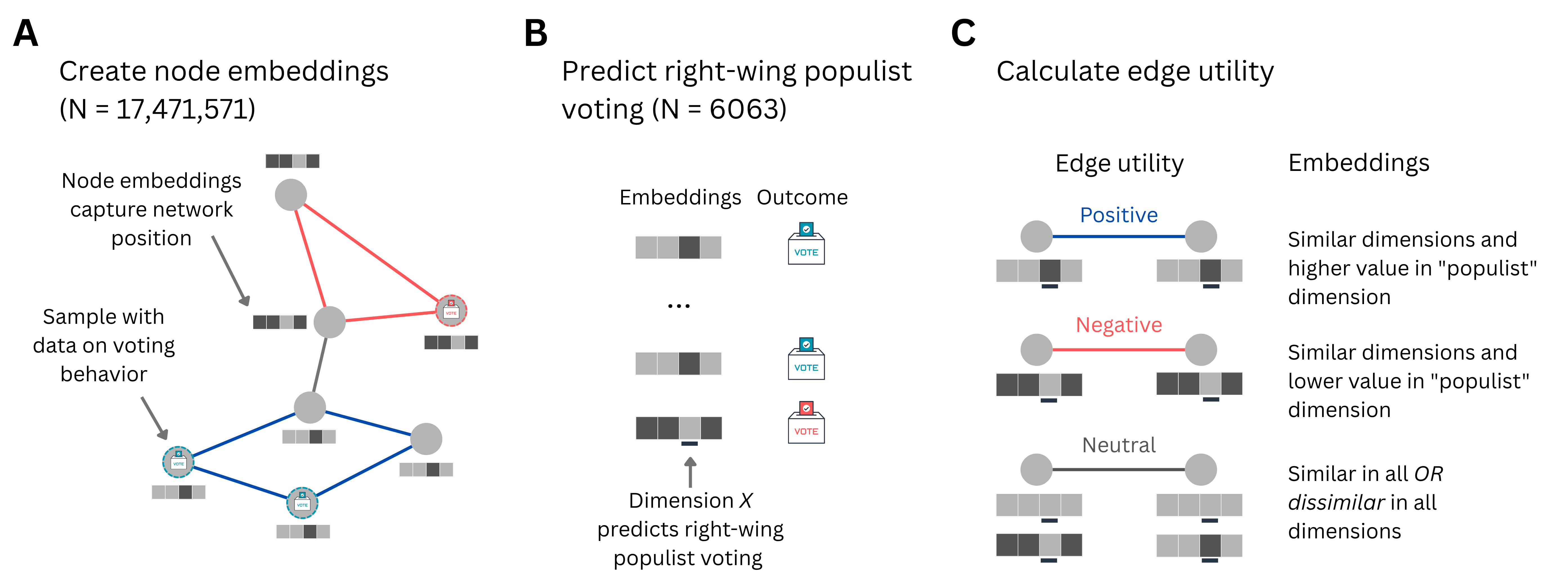}
    \caption{\textbf{Illustration of our approach}. A: We created node embeddings for all persons in the population network (here displayed with four dimensions). Persons that could be linked to their voting behavior are indicated with a dashed contour. B: For those individuals, we predicted right-wing populist voting using network embeddings and individual characteristics. We identified one dimension (underlined) that was highly predictive of right-wing populist voting. C: We computed edge utilities for the entire population network, indicating the importance of an edge for predicting right-wing populist voting. When two connected nodes have similar embedding dimensions and a higher value in the populist dimension than in the other dimensions, edge utility is positive. When they have similar embedding dimensions and a relatively lower value in the populist dimension, edge utility is negative. When they have similar embedding dimensions and the populist dimension has a similar value to the other dimensions \textit{or} completely dissimilar embedding dimensions, edge utility is zero.}
    \label{fig:workflow}
\end{figure}

While node embeddings enable us to predict right-wing populist voting from a person's network position, they are not inherently interpretable: Embedding dimensions are often highly correlated, and there is no guarantee that dimensions of the embedding space correspond to meaningful, observable external variables or social outcomes \cite{locatello_challenging_2019}. This makes it difficult to interpret the most predictive embedding dimensions to identify differences in network structure associated with the outcome. This challenge is addressed by the Dimensional Interpretability of Node Embeddings (DINE) framework \cite{piaggesi_dine_2024} that disentangles embedding dimensions. The DINE framework uses regularizing auto-encoders to enforce sparsity and orthogonality in the embedding dimensions, making them more distinct and easier to interpret. Moreover, the dimensions that are most predictive of the outcome can be mapped back to the population network. The mapping can be done by computing the \emph{edge utility} score \cite{piaggesi_dine_2024} that measures how much a specific embedding dimension contributes to the average similarity between two connected nodes, relative to all other dimensions.(Fig.~\ref{fig:workflow}C). By aggregating edge utility scores across different types of ties and demographic groups, DINE applied to population-scale node embeddings opens an opportunity to identify potentially novel cleavage dimensions in the network structure that are predictive of right-wing populist voting.

\subsection{The Present Study}

Using survey data and traditional network statistics, previous studies have shown support for a link between homogeneous network substructures along socio-economic cleavage dimensions that predict right-wing populist voting \cite{bornschier_how_2021, rydgren_legacy_2011, rydgren_social_2009, attewell_educational_2025, dejong_separated_2025} (but see also \cite{nieuwbeerta_crosscutting_2000}). Some of these studies have also demonstrated the additive predictive value of network features beyond individual characteristics. We extend this line of research by investigating the link between network structure measured through population-scale node embeddings and right-wing populist voting. The general research questions of our preregistered analysis were therefore: First, to what extent do node embeddings predict whether a person voted for a right-wing populist party in the Dutch parliament election on 22 November 2023? Second, do node embeddings have added predictive value beyond individual covariates?

In an exploratory analysis, we applied the DINE interpretability framework to answer the question whether node embedding dimensions differ in their predictive ability. This allowed us to investigate which structural patterns in the network were linked to right-wing populist voting, which is not possible using individual characteristics alone.

This study makes two contributions: Methodologically, it demonstrates how social scientists can interpret population-scale network embeddings. Substantively, we document education as a persistent cleavage dimension, supporting the re-alignment hypothesis between voting and societal divides \cite{bornschier_cleavage_2024}. Further, we show that homogeneous local network structure provides a link between education and right-wing populist voting. In sum, our paper demonstrates that node embeddings, coupled with interpretability analysis, can help researchers gain a deeper understanding of the link between network structure and social outcomes.

\section{Results}

\subsection{Pre-registered Analysis: Population Network Embeddings Predict Right-Wing Populist Voting}

We start by examining the extent to which embeddings predicted right-wing populist voting and whether they improved predictions based on individual characteristics (covariates). Among prediction models with three different feature sets---embeddings-only, covariates-only, and embeddings-plus-covariates---we expected embeddings-plus-covariates to perform better than embeddings-only and covariates-only, and any model including embeddings to outperform those without. We used Bayesian regression and marginal effects to compare the out-of-sample performance (measured via macro AUC) of prediction models across several manipulations: feature sets, prediction algorithms (k-nearest neighbors, logistic regression, and XGBoost), network years (2020, 2021, and 2022), embedding methods (DeepWalk and LINE), embedding hyperparameters, and untransformed vs. DINE-transformed embeddings. Thus, all reported results are based on macro AUC scores \textit{estimated} by the Bayesian regression. This approach allowed us to compute differences in prediction performance between feature sets aggregated over other manipulations.

Embeddings-only models showed the worst prediction performance, followed by embeddings-plus-covariates models (Fig.~\ref{fig:prediction-performance}). Covariates-only models had the highest prediction performance. The difference between embeddings-only and covariates-only was $\Delta \text{AUC}=0.09$, 95\% CI $[0.089, 0.090]$ while the difference between covariates-only and embeddings-plus-covariates models, the difference was $\Delta \text{AUC}=-0.012$, 95\% CI $[-0.013, -0.012]$. As we will explore next, this is due to the inability of logistic regression and k-nearest neighbor models to effectively use the embeddings for prediction.

Among the prediction algorithms used, XGBoost outperformed logistic regression and k-nearest neighbors (see Supplementary Table S4). Among the embedding methods and embedding hyperparameters, DeepWalk with 100 walks per node of length 20 and 32 embedding dimensions performed best; however, the differences in prediction performance were small. We did not observe differences in prediction performance between DINE-transformed and untransformed embeddings, embedding window sizes, or network years.

\subsection{Exploratory Analysis: Structural Network Differences in Education Related to Right-Wing Populist Voting}

\begin{figure}[htbp!]
    \centering
    \floatpagestyle{plain}
    \includegraphics[width=0.8\textwidth]{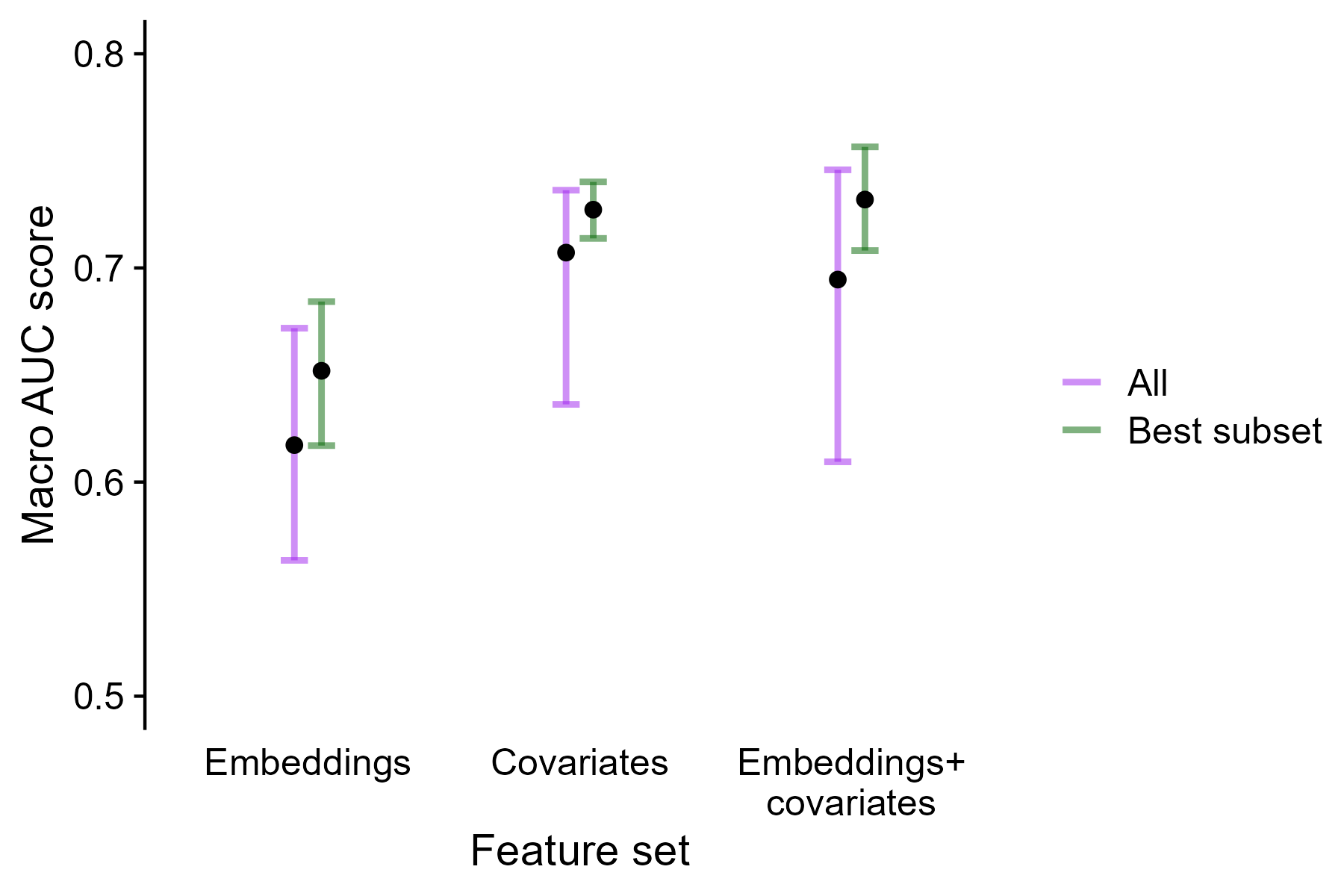}
    \caption{\textbf{Out-of-sample prediction performance for right-wing populist voting}. Performance was measured with the macro AUC score (on y-axis). The x-axis shows different feature sets used for prediction. Points indicate the posterior predictive mean. Vertical bars indicate 95\% credible intervals. Colors indicate whether the predictions were made for all scores (purple) or only for a subset of the best performing prediction models (XGBoost and embeddings with 100 walks per node of length 20 and 32 embedding dimensions, green).}
    \label{fig:prediction-performance}
\end{figure}

Because some embedding hyperparameters and prediction algorithms performed better than others, we compared prediction performance across feature sets within the best subset of predictive models, namely, XGBoost using DeepWalk embeddings with 100 walks of length 20 with dimension 32. Aggregated over manipulations that did not show any difference in performance (untransformed vs. DINE-transformed embeddings, window sizes, and network years), embeddings-plus-covariates models were slightly better than covariates-only models, $\Delta \text{AUC} = 0.011$, 95\% CI $[0.009, 0.012]$ (Fig.~\ref{fig:prediction-performance}).

For the remainder of our exploratory analysis, we selected the DINE-transformed embeddings from the prediction model with the highest prediction performance score (for the prediction model and embedding configuration, see Supplementary Table S7). 

\subsubsection{One Embedding Dimension Is Strongly Associated With Right-Wing Populist Voting}

After observing that network embeddings predicted right-wing populist voting, we investigated whether embedding dimensions differed in their importance for prediction. The regularizing auto-encoder in the DINE framework transforms embedding dimensions to become orthogonal and sparse (see Methods for details). We therefore expected the feature importance of DINE-transformed embeddings to be concentrated in fewer dimensions.

\begin{figure}[htbp]
    \centering
    \floatpagestyle{plain}
    \includegraphics[width=0.8\textwidth]{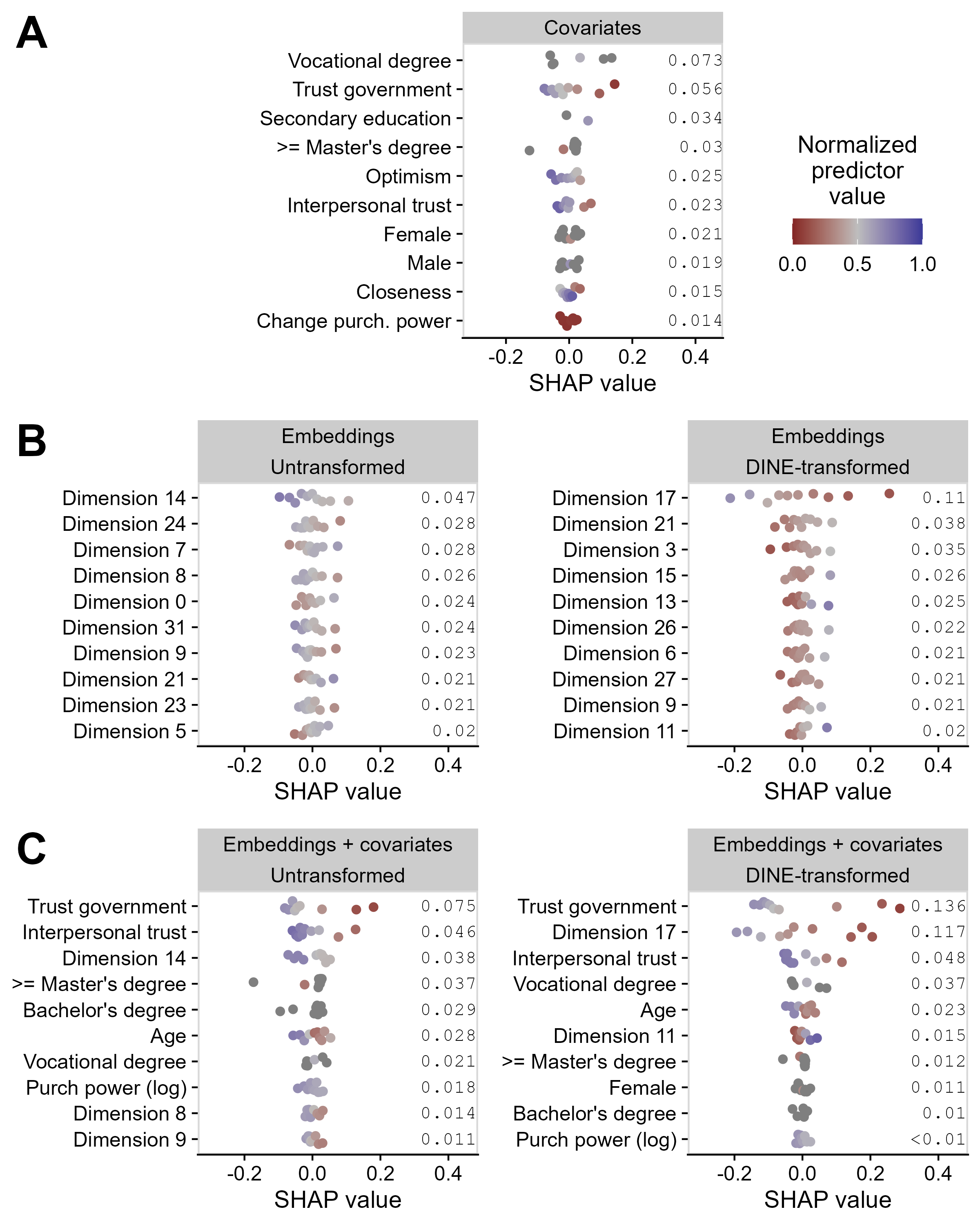}
    \caption{\textbf{Importance of 10 most important variables predicting populist voting behavior in the LISS panel data}. Importance was quantified with SHAP values for each predictor and observation. Individual SHAP values were aggregated for each decile to guarantee privacy of the panel subjects. Each point represents the average SHAP value of a decile. Color indicates the average SHAP value in the decile, normalized between zero and one. Grey-colored points indicate that the average value could not be published to prevent group disclosure. Predictors on the y-axis are ordered according to the mean of their absolute decile-averaged SHAP values which are displayed on the right side of each panel. Panels contain results for predictor sets that included (A) only covariates (B) only embeddings or (C) embeddings and covariates. For (B--C), the results with untransformed versus DINE-transformed embeddings is shown in the two columns.}
    \label{fig:feature-importance}
\end{figure}

To assess the effect of the DINE-transformation and of combining embeddings with covariates, we used SHAP values to measure the contribution of each feature to the model’s predictions across the three feature sets (covariates-only, embeddings-only, embeddings-plus-covariates), both when the embeddings were untransformed and when they were transformed using DINE (Fig.~\ref{fig:feature-importance}).

Starting with a prediction model using only covariates (Fig.~\ref{fig:feature-importance}A), we observed that the highest achieved education levels and trust in the government were the most important predictors, followed by optimism, interpersonal trust, and gender. Highest achieved vocational education, lower trust in the government, lower optimism, and lower interpersonal trust increased the predicted probability of voting for a right-wing populist party.

We also compared embeddings-only prediction models (Fig.~\ref{fig:feature-importance}B). As expected, we observed that the model with untransformed embeddings had SHAP values evenly distributed across dimensions. This suggests that they were roughly equally important for predicting populist voting. For models with DINE-transformed embeddings, SHAP values were larger for dimension 17 than for other dimensions, indicating that this dimension was more important than others.

Next, we compared models that included both embeddings and covariates (Fig.~\ref{fig:feature-importance}C). In the model with untransformed embeddings, multiple embedding dimensions had relatively small SHAP values. In the model with DINE-transformed embeddings, dimension 17 had much larger values than the other dimensions, being the second most important predictor after trust in the government.

\subsubsection{Structural Differences in ``Populist" Embedding Dimension}

\paragraph{Individual Level}

Because dimension 17 was the DINE-transformed embedding dimension most predictive of right-wing populist voting, we selected it to calculate utility scores for all edges in the population network. Edge utility is a continuous measure indicating whether dimension 17 contributes to the similarity between two connected
nodes, relative to all other dimensions (i.e., it captures the \textit{marginal} contribution to the average element-wise product between source and target embeddings; see Methods for details). In our interpretation, edge utility is positive when two persons have similar node embeddings and above-average values in dimension 17, suggesting that they were less likely to vote right-wing populist. Conversely, edge utility is negative when two persons have similar node embeddings and below-average values in dimension 17, suggesting that they were more likely to vote right-wing populist. Taken together, edge utility in dimension 17 reflects the importance of an edge for predicting right-wing populist voting.

\begin{figure}[htbp!]
    \centering
    \floatpagestyle{plain}
    \includegraphics[width=1\textwidth]{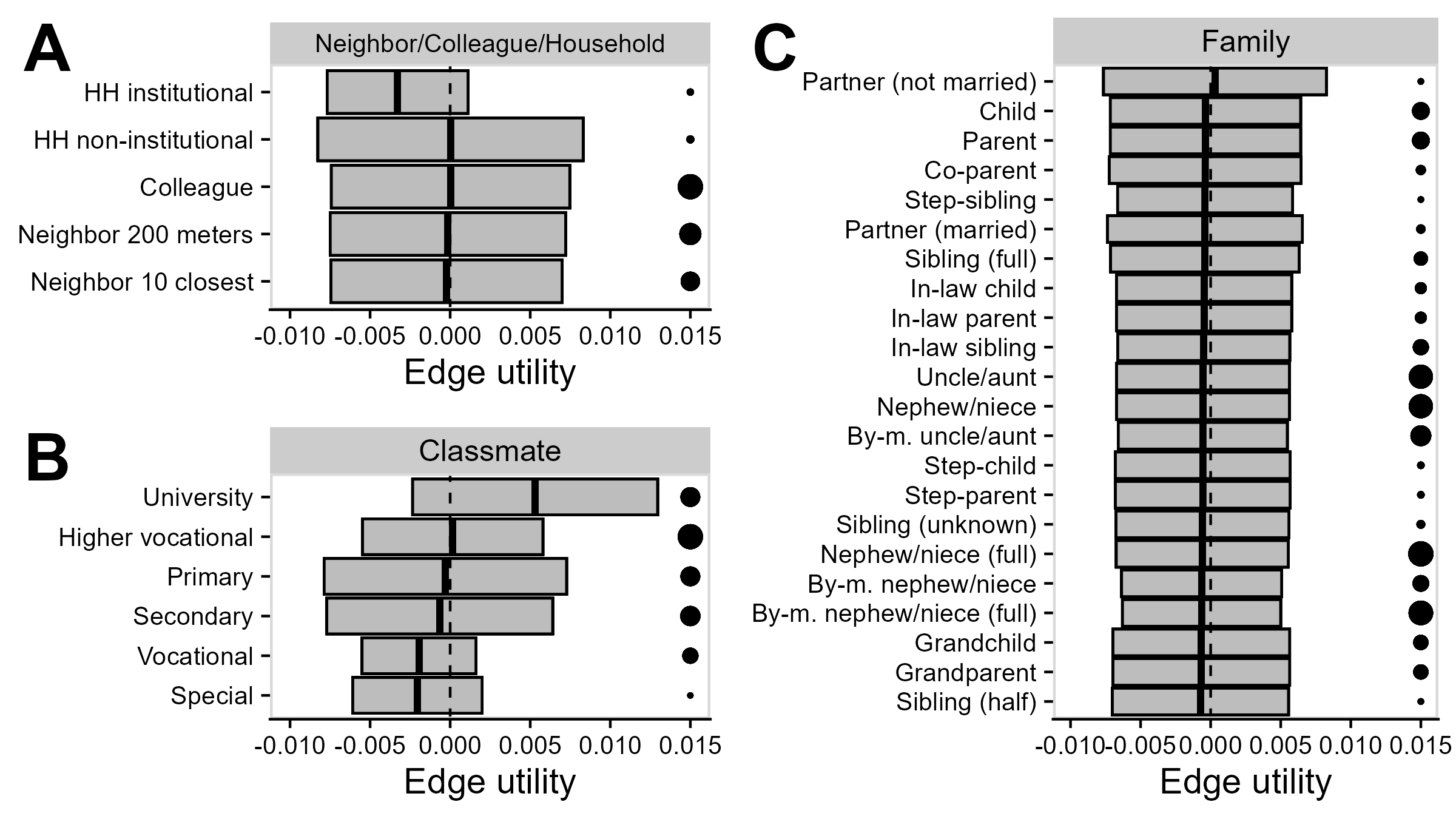}
    \caption{\textbf{Edge utility scores (x-axis) for different network relation types (y-axis) and relation groups}. Bold vertical bars indicate means and boxes span $\pm$ 2 standard deviations from the mean of the edge utility distributions of (A) neighbors, colleagues and households, (B) classmates, and (C) family relations. The size of the circles on the right of each box indicate the number of relations of the respective type. Classmate relations typically refer to students enrolled at the same school in the same location in the same school year (but this differs per relation type). Special schools are dedicated to students with mental, physical, or learning disabilities. Examples for institutional households are elderly homes, student dorms, and prisons. For neighbor relations, ``200 meters'' refers to 20 random persons living within a 200 meter radius and ``10 closest'' refers to all persons living at the 10 closest addresses. HH = household. By-m. = related by marriage.}
    \label{fig:edge-covariates}
\end{figure}

We compared edge utility between different relation types in the population network. Most relation types had an edge utility distribution centered around zero (Fig.~\ref{fig:edge-covariates}) suggesting little systematic importance of dimension 17. However, institutional household relations showed a negatively shifted edge utility distribution compared to non-institutional households and other relation types (Fig.~\ref{fig:edge-covariates}A). Among classmate relations, the edge utility distribution was positively shifted for university relations, and negatively---but less strongly---shifted for vocational and specials school relations (Fig.~\ref{fig:edge-covariates}B). For primary, secondary, and higher vocational school relations, edge utility distributions were centered around zero. This suggests that there were differences in network structure between university and vocational classmates that were predictive of right-wing populist voting: Both university and vocational students were connected to similar others, but the former were less likely, and the latter more likely, to vote populist.

\begin{figure}[htbp!]
    \centering
    \floatpagestyle{plain}
    \includegraphics[width=1\textwidth]{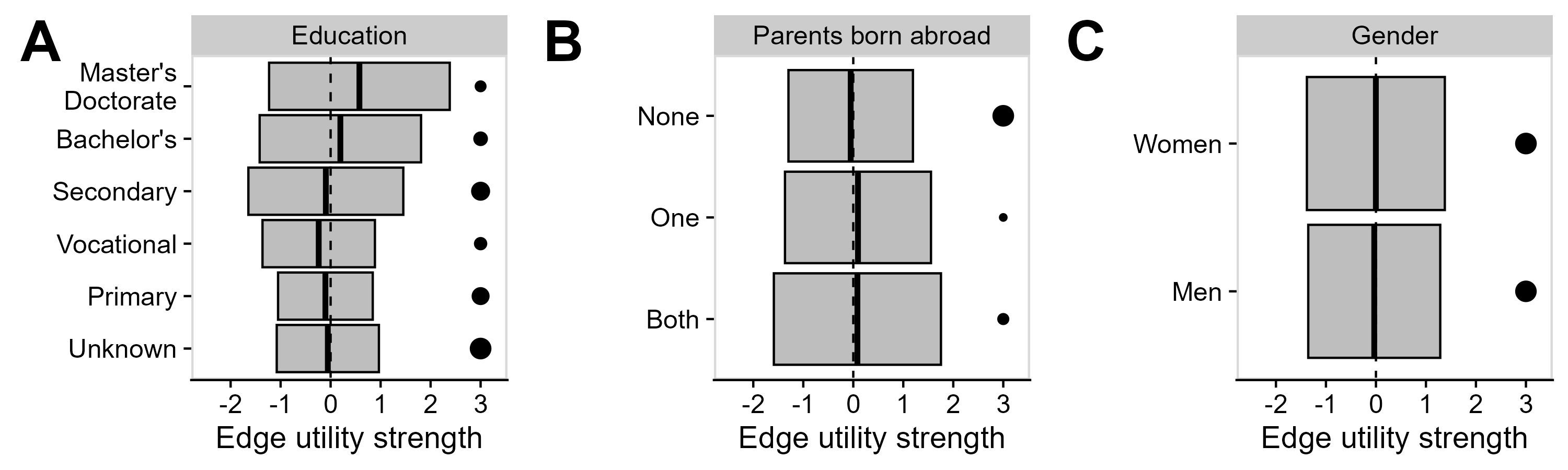}
    \caption{\textbf{Edge utility strength for different person-level variables}. Highest achieved education levels (A), the number of parents born outside the Netherlands (B), and gender (C). Bold vertical bars indicate means and boxes span $\pm$ 2 standard deviations from the mean of the edge utility strength distributions. The size of the circles on the right of each box indicate the number of persons in the respective category. The categories are mutually exclusive within each panel but not across panels.}
    \label{fig:node-covariates}
\end{figure}

Next, we looked at the importance of relations in each person's ego-network measured by edge utility strength, i.e., the sum of the edge utility in dimension 17 from first-order relations (Fig.~\ref{fig:node-covariates}). Edge utility strength measures the structural similarity between a person and its first-order neighbors that is associated with right-wing populist voting. We observed differences in edge utility strength between achieved education levels (Fig.~\ref{fig:node-covariates}A) but no differences for migration background (Fig.~\ref{fig:node-covariates}B) or registered gender (Fig.~\ref{fig:node-covariates}C). For achieved education, the results show that persons who obtained a Master's/doctorate or Bachelor's degree had on positively shifted, and those with a vocational degree had negatively shifted utility strength distributions. Persons who had achieved a primary, secondary, or unknown education level had edge utility distributions centered close to zero. This pattern indicates that persons with university and vocational achieved education levels were more likely to be connected to others with similar voting behavior. For those who obtained a Master's or doctorate degree this effect was stronger than for those who obtained a Bachelor's or vocational degree. This supports our earlier finding that there were homogeneous local network substructures that were predictive of right-wing populist voting and captured by dimension 17.

We also looked at Pearson correlations between edge utility strength and age as well as gross income percentile. Both correlations were relatively small (age: $r=-0.05$; income: $r=0.08$), suggesting little difference in structural similarity in dimension 17 with respect to age and income.

\paragraph{Municipality Level}

\begin{figure}[htbp!]
    \centering
    \floatpagestyle{plain}
    \begin{subfigure}{\textwidth}
        \includegraphics[width=1\textwidth]{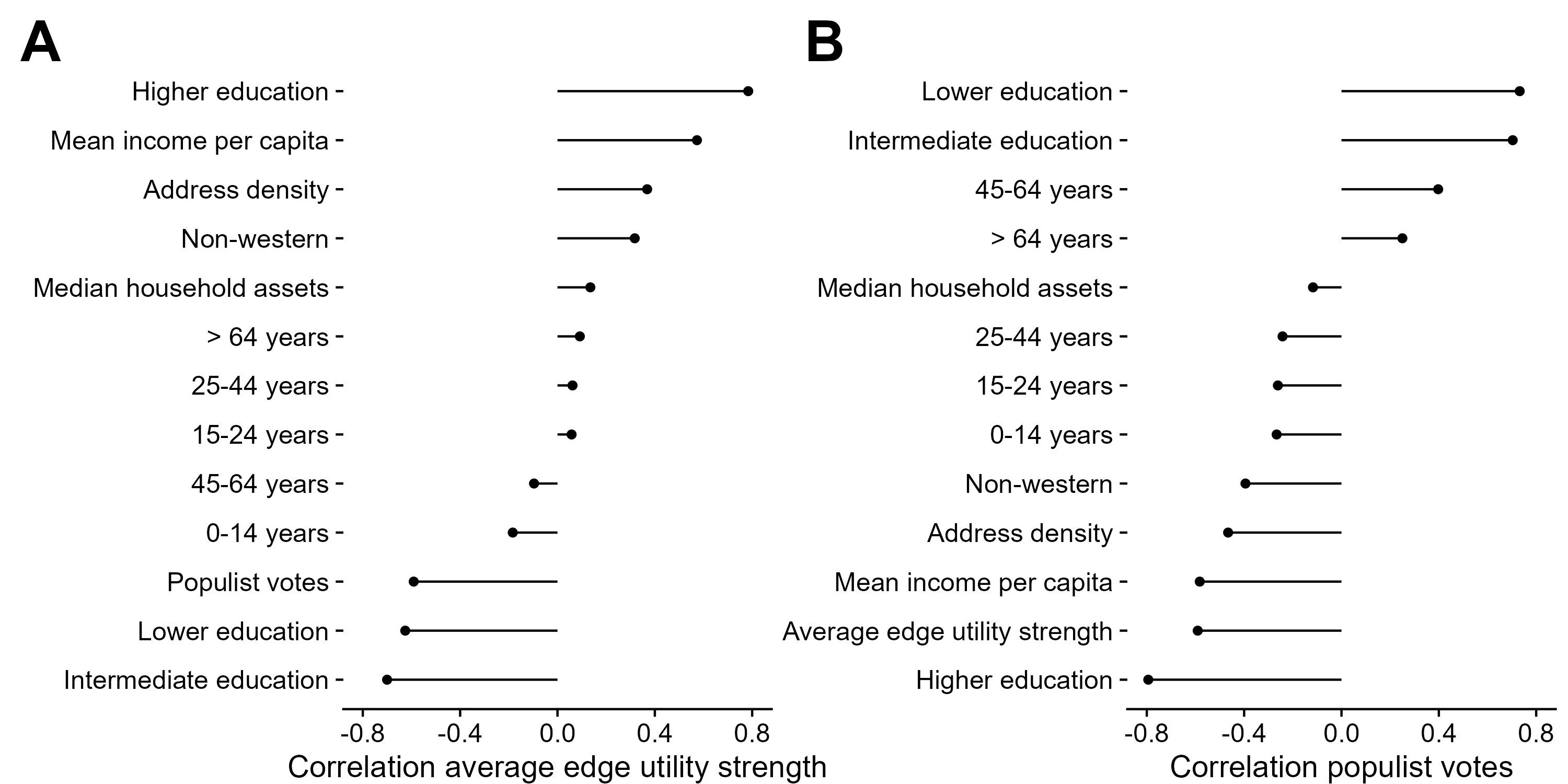}
    \end{subfigure}
    \par\medskip
    \begin{subfigure}{\textwidth}
        \includegraphics[width=1\textwidth]{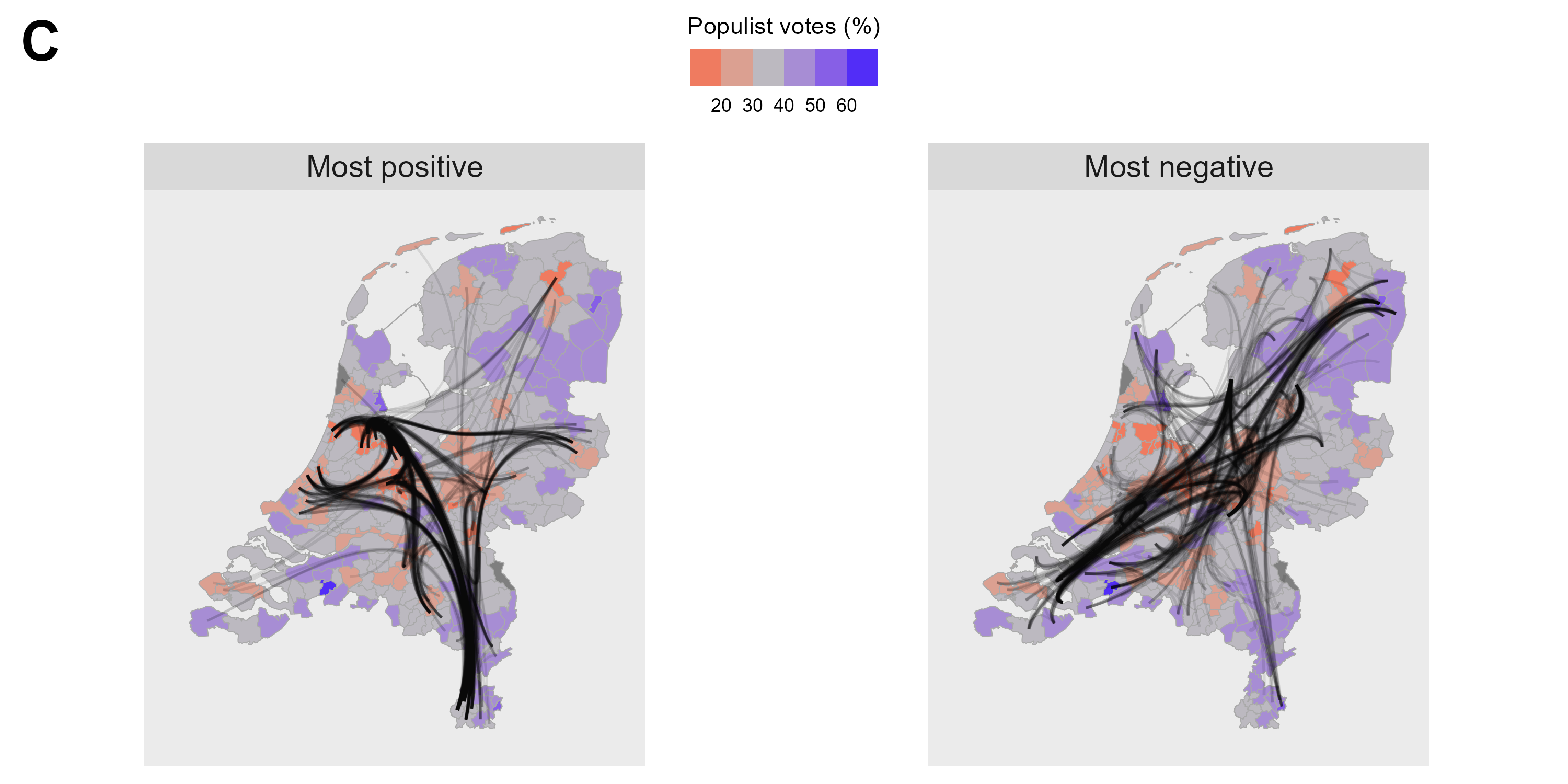}
    \end{subfigure}
    \caption{\textbf{Average edge utility strength at the municipality level}. Pearson correlations between aggregated municipality statistics and (A) municipality average edge utility strength and (B) right-wing populist votes. C: Maps of the Netherlands and average edge utility connections between Dutch provinces. The left panel shows the most positive connections (top percentile of average edge utility distribution) and the right panel shows the most negative connections (bottom percentile). Color indicates the percentage of votes for a right-wing populist party in each province (gray represents the country-wide average).}
    \label{fig:municipality}
\end{figure}

Because prior research has shown differences in the dynamics of populist voting behavior at the micro- and macro-level \cite{berman_causes_2021}, we extended our analysis to the municipality-level. Moreover, aggregated voting results were publicly available at the municipality level, allowing us to link aggregated edge utility and right-wing populist voting directly. That is, we computed the average edge utility between persons residing in the same or in a different municipality in the Netherlands. For each municipality, we then calculated the strength of the average edge utility. The strength reflects the average importance of persons' relations for predicting right-wing populist voting in a municipality. We correlated the average edge utility strength with municipality-level statistics, including right-wing populist votes. Figure~\ref{fig:municipality}A shows that municipalities with more university-educated inhabitants, higher income, higher address density (i.e., more urban), and more inhabitants with non-western migration background had higher average edge utility strength, suggesting higher importance for right-wing populist voting. Thus, residents in higher education, income, and address density municipalities were similar in their network position and less likely to vote right-wing populist, whereas residents in vocational education municipalities were also similar in their network position but more likely to vote right-wing populist. Age groups, and household assets had low correlations with average edge utility strength and, thus, were less important at the municipality level.

Figure~\ref{fig:municipality}B shows correlations between municipality statistics and right-wing populist votes. Education levels had the strongest correlations ($|r| \approx 0.8$), followed by average edge utility strength and income($|r| \approx 0.6$). Thus, average edge utility strength was very strongly associated with right-wing populist voting at the municipality level, indicating that municipalities with positive average edge utility strength had fewer and municipalities with negative average edge utility strength had more populist votes.

To provide a geographical intuition on which inter-municipality relations were the most important for predicting right-wing populist voting, we show maps of Dutch municipalities with the most positive ($>99$th percentile) and most negative ($<1$st percentile) average edge utilities between them in Fig.~\ref{fig:municipality}C. The most positive average edge utilities were between municipalities with larger cities, such as Amsterdam, Rotterdam, and Maastricht, which also had low percentages of right-wing populist votes. The most negative average edge utilities were between rural, smaller municipalities that range from the North-east (e.g., in the province of Drenthe) to the South-west (e.g., in the province of Zeeland). These municipalities tended to have high percentages of right-wing populist votes. Thus, relations between persons living in urban as well as remote rural municipalities were the most important for predicting right-wing populist voting.

\subsubsection{Sensitivity Analysis}

Both embedding methods that we used rely on random sampling. To see how much our exploratory results depended on the specific embeddings we selected, we repeated the analysis with the DINE-transformed embeddings from the second-best prediction model (see Supplementary Table S7 for model and embedding hyperparameters and Supplementary Materials for results). The results were overall similar: One transformed embedding dimension was strongly associated with right-wing populist voting (although it was only the third most important predictor after trust in the government and vocational degree). Analyzing edge utility in this dimension revealed similar differences in edge utility between household and classmate relations and no meaningful differences between family relations. Similar to the best model, edge utility strength was on average positive for university and negative for vocational education levels. However, the correlation between edge utility strength and income was higher ($r=0.17$) than in the main analysis ($r=0.08$), suggesting that persons with higher income had more important first-order relations and structural network differences compared to persons with lower income. At the municipality level, correlations between average edge utility strength and municipality statistics were overall lower but similar in rank.

\section{Discussion}

In this study, we created node embeddings based on formal admninistrative ties reflecting shared social contexts for the entire Dutch population network. We used these embeddings to predict right-wing populist voting in a representative sample. In the preregistered part of the study, we show that, in line with our expectations, embeddings predicted right-wing populist voting above chance level. However, in contrast to what we expected, predictions based on embeddings alone were worse than those based on personality-related and socio-economic covariates, and the embeddings did not improve covariate-based predictions. Nevertheless, when looking at a subset of the best-performing embeddings in our exploratory analysis, we found that embeddings slightly improved the prediction performance and that one embedding dimension was among the most important predictors.

The hyperparameters of our embedding methods (e.g., random walk length, Skip-gram window size, number of dimensions) had little or no influence on our prediction results compared to the type of prediction algorithm and the set of predictors (see Supplementatry Table S4). This means that close neighbors were likely more important for predicting right-wing populist voting compared to more remotely connected others and that the predictive network structure can be summarized with few latent dimensions.

In the Supplementary Materials, we also present prediction results for two additional outcomes: general voting behavior and trust in the government. They show similar performance patterns, indicating that the predictive ability of network embeddings generalizes to voting beyond right-wing populist parties and political attitudes related to populist voting \cite{colloca_rally_2024}. For general voting, the increase in prediction performance from combining embeddings and covariates was slightly larger compared to right-wing populist voting when looking at the best subset of embeddings.

Our study showed that, for social ties based on administrative data, the information encoded in the network position was almost entirely accounted for by established individual covariates of right-wing populist voting. Our exploratory analysis gives insight into which covariates overlapped with our embeddings. One embedding dimension was particularly important for predicting right-wing populist voting after transforming the embeddings using the DINE framework \cite{piaggesi_dine_2024}. In principle, DINE allows researchers to uncover dimensions different from observable socio-demographics that separate a network into different (local) substructures and are predictive of a target outcome. In our study, we found that the most predictive embedding dimension was highly aligned with current and achieved education at the individual level and municipality level. That is, this ``populist'' embedding dimension captured differences in local network structure between university-level and vocational education that were predictive of right-wing populist voting. Given that node embeddings did not provide additional explanatory value, one might ask whether network position merely proxies for achieved education. However, when included in the same prediction model, both education level and network position retained predictive power, indicating that they capture related but non-redundant information.

Previous studies have found support for education as a social cleavage dimension that is aligned with homogeneous network substructures that facilitate the embrace of right-wing populist ideologies \cite{attewell_educational_2025, dejong_separated_2025} (but see also \cite{nieuwbeerta_crosscutting_2000}). While these studies have used surveys generating small ego-networks from informal ties and traditional network statistics, we provide additional evidence for this relationship based on population-scale formal ties and node embeddings. Our findings support the notion of education as an important cleavage dimension that predicts and potentially shapes voting behavior and evidence in favor of the re-alignment between voting and societal divides \cite{bornschier_cleavage_2024}. Furthermore, our results suggest that university-educated persons were ``embedded'' in more homogeneous local network structures than vocationally educated persons. This finding is in line with the mass society hypothesis and supports previous evidence on asymmetries between educational groups and populist voting \cite{vannoord_educationbased_2025}.

In contrast to previous studies (e.g.,\cite{attewell_educational_2025, dejong_separated_2025}), we did not find substantial added predictive value of features encoding network structure for right-wing populist voting. This limitation likely stems from the construction of the embeddings: The network layers differ qualitatively in the probability of resulting in actual interactions between persons. For example, living in the same household is more likely to result in frequent interaction than living in the same neighborhood. Regarding voting behavior and related contagious social outcomes, different layers contain predictive information due to homophily, contagion, and shared environments \cite{shalizi_homophily_2011}. By treating the layers equally when creating the embeddings, we likely overshadowed contagion with homophily and shared environment effects that were already captured by individual characteristics included in our analysis.

Do our findings provide insight into the causal mechanisms between education, network position, and right-wing populist voting? A recent study using data from the LISS panel has shown that adding or losing a close informal tie to someone with the same education level predicted changes in voting behavior \cite{dejong_separated_2025}. This supports the interpretation that education and network position shape voting behavior and not \textit{vice versa}. Furthermore, we found that university ties were highly homogeneous with respect to populist voting compared to primary and secondary school relations. This suggests a crucial role of higher education as a causal driver for the creation of homogeneous network structures that persist into post-educational life trajectories \cite{hooghe_field_2025}. However, given that our analysis did not account for pre-school socialization and selection processes, this interpretation must be handled with care.

Our study comes with the limitation that registry-based population networks in their current state miss important shared social contexts, such as voluntary associations (e.g., religious groups), that have high probabilities of resulting in frequent interactions \cite{putnam_bowling_2000}. Thus, future studies could explore linking registry networks with other network data sources, for example, donated social media data or representative surveys, and explore their added predictive value \cite{sakshaug_augmenting_2023}. 

While our embeddings encoded the network position of the entire Dutch population, we only predicted retrospective voting behavior measured in a survey administered to the much smaller LISS panel. The LISS panel aims to be representative of the Dutch population (but see \cite{knoef_representativeness_2009}), and we thus expect our prediction results to generalize to the population-level. Furthermore, we observed similar if not stronger results at the municipality level, supporting the generalizability of the individual-level results. However, we faced the problem that not all participants in the LISS panel could be linked to the registry data, for example, because they opted out of the linkage. While this could potentially affect the generalizability of our prediction results \cite{sakshaug_augmenting_2023}, we did not observe substantial differences in right-wing populist voting before and after the linkage (17.8\% vs. 17.9\%). Another, more severe issue was that fewer people in the LISS panel indicated that they voted for a right-wing populist party compared to official voting results at the national level (31.1\% votes for right-wing populist parties). We explain this discrepancy with unexpected voter shifts from undecided and non-populist parties towards the right-wing populist PVV \cite{vanholsteyn_dutch_2025}. Given the relatively small sample size of the LISS panel, it might not be able to reflect such strong electoral volatility. Two alternative explanations are that the LISS panel differs from population statistics in characteristics that are associated with right-wing populist voting (e.g., post-secondary education, non-western migration background; \cite{knoef_representativeness_2009}) and that populist voters are less likely to enter a panel, respond in questionnaires, or reveal their true voting intentions because they have less trust in institutions \cite{scherpenzeel_true_2010}. The discrepancy between survey and official voting outcomes might also explain why we observed stronger associations at municipality compared to the individual level.

Using social media networks, previous studies have employed parametric latent space models to infer ideological stance from user interactions by aligning estimated latent dimensions with survey responses \cite{barbera_birds_2015, ramaciottimorales_inferring_2022}. Due to high alignment between estimates and survey responses, these ideological stances could be reliably extrapolated beyond the surveyed subjects. Our use of DINE with subsequent analysis of edge utility in the ``populist'' dimension can be understood as an analogous interpretability strategy, operating at the edge level rather than the node level: instead of aligning node positions with expert placements \cite{barbera_birds_2015, ramaciottimorales_inferring_2022}, we identified the most predictive dimension and interpreted it by examining how edges in that dimension related to individual-level survey data (voting behavior) and administrative records (education level). Although our population embeddings predicted right-wing populist voting above chance level, prediction performance was not sufficient to impute individual voting behavior across the entire Dutch population. Nevertheless, the ``populist'' embedding dimension emerged as the most important predictor available at population scale (Fig.~\ref{fig:feature-importance}), suggesting that it can serve as an efficient indicator of whether a person was more or less likely to vote for a right-wing populist party.

An important difference between our setting and the latent space models used for ideology estimation from social media concerns the nature of the network ties. Social media follow or retweet networks are \emph{choice-based}: users actively select whom to follow, so tie formation directly reflects ideological affinity---the core assumption underlying spatial following models. The administrative population network we use is \emph{context-based}: ties arise from shared institutional settings (neighborhoods, workplaces, schools, households) rather than voluntary ideological affiliation. This means that the latent dimensions we recover capture structural positions shaped by institutional sorting processes rather than expressed political preferences, which may explain why our embeddings are correlated with observed socio-demographic covariates. 

We intended the embeddings to be generic individual representations of shared social contexts. We see the embeddings used in this study as a starting point and encourage future studies to test more outcome-specific embeddings, more complex embedding methods (e.g., graph neural networks
\cite{corso_graph_2024}), or layer-specific embeddings (e.g., layered random walks \cite{liu_principled_2017}). Alternatively, future embeddings could be created from a subset of contexts (e.g., only close family relations in the family layer) or shared contexts defined by different rules. While we did not find differences in the predictive performance of embeddings created for years 2020 to 2022, we think it is worthwhile to explore changes in network structure through embeddings over a longer period, extending previous findings from the Dutch and Danish population network \cite{cremers_unveiling_2024, kazmina_can_2025, bokanyi_fragmentation_2026}.

Before concluding this article, we want to emphasize that the predictive models in this study were designed as proof-of-concept to show how much predictive information population network embeddings contain compared to other variables and to identify the most important embedding dimensions. They are not intended for deployment, for example, to predict voting outcomes in future elections in the Netherlands. Given that elections can be influenced by political history (e.g., passing certain laws) and external events (e.g., wars) that are not necessarily reflected in the data we used for prediction, we do not know the extent to which our models generalize beyond the election used for training and evaluation.

The conclusion from this study is that educational divides are an important predictor and potential causal driver of right-wing populist voting behavior. Going beyond small ego-networks obtained from surveys and handcrafted statistics, interpretable embeddings that automatically capture the structure of a population-scale network allowed us to identify homogeneity in local network structure as a potential explanation for this association. We believe that population-scale embeddings and interpretability methods such as DINE can be fruitful to study other social outcomes that are more difficult to predict with individual characteristics.


\section{Methods}\label{sec:methods}

We preregistered the methodology and analysis plan of the first part of the study on the Open Science Framework (\url{https://osf.io/td68b/}). The repository also contains aggregated data that were approved for publication by Statistics Netherlands (see also our Ethics statement). We report deviations from the preregistration in the Supplementary Materials using a template by \cite{willroth_best_2024} and assess their impact on the test severity and inference quality of our preregistered analysis \cite{lakens_when_2024}. A more detailed explanation of datasets and variables with identifiers can also be found in the Supplementary Materials.

\subsection{Data}

\subsubsection{Registry Data}

We created node embeddings based on the Dutch population network available through Statistics Netherlands. The nodes in the network represent persons registered in the Netherlands. We used ties from all five available network layers that reflect shared social contexts between persons: A neighbors layer that links individuals to the persons living at the geographically 10 closest addresses (not considering height differences or institutional households with more than 10 members) as well as persons living at one of 20 addresses within 200 meters (the latter are randomly sampled when more than 20 households exist within 200 meters and include institutional households) \cite{centraalbureauvoorstatistiek_burennetwerk_2009}. A colleagues layer connecting a person to the 100 geographically closest persons working for the same employer \cite{centraalbureauvoorstatistiek_colleganetwerk_2009}. A family layer connecting persons with their nuclear (parents and children) and extended family (grandparents, cousins, nephews, nieces, aunts, and uncles) \cite{centraalbureauvoorstatistiek_familienetwerk_2009}. A household layer connecting all persons living in the same household \cite{centraalbureauvoorstatistiek_huisgenotennetwerk_2009}. And a classmates layer which connects persons enrolled in the same institution, institution location, institution type, or educational program (e.g., primary school, university), and study year \cite{centraalbureauvoorstatistiek_klasgenotennetwerk_2009}. Details on the network creation from registry data, basic network statistics, and data access are described in \cite{vanderlaan_whole_2023, bokanyi_anatomy_2023}. While the neighbors and colleagues layers are not necessarily symmetric, we treated them as such by adding reverse connections for each directed connection and filtering for unique connections. We collapsed the five layers into a single undirected, unweighted network (also referred to as \textit{network aggregation} \cite{liu_principled_2017}). We justify creating an unweighted network by the low number of persons that are connected in more than one layer (see \cite{bokanyi_anatomy_2023} for layer overlap in the 2018 population network). We created collapsed networks for the years 2020, 2021, and 2022 to allow for comparisons across years.

Besides the network, we used several person-level variables that are available in the registry: Basic demographic variables were age (years since birth year), gender (male, female, unknown), and number of parents born outside of the Netherlands (none, one, or both, unknown). Socio-economic variables were $\text{log}(1+x)$-transformed purchasing power (i.e., average of income of current and previous year standardized according to prices of current year), change in purchasing power from pervious to current year, gross income percentile, and highest achieved educational degree based on aggregation level two of the \textit{Standaard Onderwijsindeling 2021} by Statistics Netherlands (primary, vocational, secondary, Bachelor's or equivalent, Master's or doctorate; see \url{https://www.cbs.nl/nl-nl/onze-diensten/methoden/classificaties/onderwijs-en-beroepen/standaard-onderwijsindeling--soi--/standaard-onderwijsindeling-2021}).

\subsubsection{Survey Data}

In addition to the registry data, we make use of data from the LISS panel (Longitudinal Internet studies for the Social Sciences; \cite{scherpenzeel_true_2010}; accessed on 17 September 2024), which is administered and managed by the non-profit research institute Centerdata (Tilburg University, Netherlands). The LISS panel is a representative sample of Dutch individuals who participate in monthly internet surveys. The panel is based on a true probability sample of households drawn from the population register by Statistics Netherlands. Self-registration is not possible, and households that would otherwise be unable to participate are provided with a computer and internet connection.

We used data from two modules of the LISS Core Study wave 16 collected between 2023 and 2024. From the Personality module \cite{marchand_liss_2024}, we used the variables happiness, trait mood, state mood, life satisfaction, interpersonal trust, extraversion, emotional stability (i.e., inverse of neuroticism), openness, conscientiousness, agreeableness, self-esteem, interpersonal closeness, social desirability, and optimism.

We created three outcome variables from the Personality and Values module \cite{elshout_liss_2024}: The first two were trust in the government (``Can you indicate, on a scale from 0 to 10, how much confidence you personally have in each of the following institutions?'' with the item ``The Dutch government'') and general voting behavior (``For which party did you vote in the parliamentary elections of 22 November 2023?''). Panel participants could choose from a list of 17 national parties, and we added additional categories for missing answers and non-voters. From this variable, we constructed the third outcome ``voted for right-wing populist party'' by categorizing answers according to (non-) right-wing populist parties, missing, not eligible, or not voted. We used the Populist 3.0 \cite{rooduijn_populist_2024} to categorize parties as (non-) right-wing populist. Because version 3.0 of the populist was published in 2022, we classified parties that were not considered in the Populist as non-populist. Since the only left-wing political party classified as populist is the Socialist Party, and it is considered a ``borderline populist" party, we decided to exclude it and focus only on right-wing populist parties (i.e., PVV, BBB, Ja21, FvD). For conciseness, we focus on the outcome right-wing populist voting in the main text and report results on the other two outcomes in the Supplementary Material and refer to them in the discussion.

We used linkage keys provided by Statistics Netherlands to link the panel variables with the registry networks and person-level variables. Some panel participants opted out of the linkage or could not be linked for unknown reasons (13.9\%), resulting in a linked panel sample size of 6063 participants. Descriptive statistics of the linked panel and person-level registry variables are provided in the Supplementary Materials.

We removed 136 participants who were not eligible for voting (2.24\%; 2.25\% before data linkage).

\subsection{Network Embeddings}

\subsubsection{Embedding Generation}
For the undirected, unweighted population network $\mathcal{G}(\mathcal{V,E})$, we created an embedding matrix $\mathbf{X} \in \mathbb{R}^{D \times  |\mathcal{V}|}$ with $D$ dimensions using two different methods: First, we trained Skip-gram models \cite{mikolov_distributed_2013} that predict node pairs in random walk node sequences generated from the network. This method is known as DeepWalk \cite{perozzi_deepwalk_2014}. We trained DeepWalk models on 10 and 100 random walks of length 5, 10, and 20 starting at each node in the network using a context window size of 2, 5, and 8, and embedding dimension 32, 64, and 128. We trained models with 10 walks per node for 20 epochs, and models with 100 walks per node for 2 epochs. We used an initial learning rate of 0.025, a minimum learning rate of 0.0001, and a negative sampling ratio of 5 with an exponent of 0.75. Second, we trained LINE models that predict whether a pair of nodes is a positive sample (they have a connection in the network) or a negative sample (the nodes are randomly sampled). Our LINE models contain embeddings that encode first- and second-order node relationships. First-order embeddings encode local neighborhoods, whereas second-order embeddings capture structural similarity (i.e., nodes with a similar position in their local neighborhood have similar embeddings). We concatenate the two embedding vectors with 8, 16, 24 dimensions, respectively. We used a learning rate of 0.025 and trained for 5 epochs. The negative sampling ratio was 5 with an exponent of 0.75 and the batch size was 100,000 edges.

\subsubsection{Decoupling Embedding Dimensions to Improve Interpretability}
We further transformed the embeddings using the DINE framework \cite{piaggesi_dine_2024} to obtain orthogonal, sparse, and more interpretable embedding dimensions. 
In particular, we employed a one-layer autoencoder to learn interpretable node embeddings. For the input embedding matrix $\mathbf{X}$, the hidden representation is computed as $\mathbf{H}=\text{sigmoid}(\mathbf{W}^{(0)}\mathbf{X} +\mathbf{b}^{(0)})$ and the autoencoder reconstructs the input as $\tilde{\mathbf{X}}=\mathbf{W}^{(1)} \mathbf{H} + \mathbf{b}^{(1)}$. The hidden layer matrix $\mathbf{H} \in \mathbb{R}^{D \times  |\mathcal{V}|}$ contains the transformed embeddings that we aim to learn. While the original embeddings have range $(-\infty,\infty)$, the transformed embeddings have range $[0,1]$. 

We trained the autoencoder using the loss function:
\begin{align}
    \mathcal{L} &= \text{MSE}(\mathbf{X},\tilde{\mathbf{X}}) + \mathcal{L}_{orth} + \mathcal{L}_{size},
\end{align}
where $\mathbf{X}$ and $\tilde{\mathbf{X}}$ are the original and predicted embedding matrices, respectively. 

For each dimension $d = 1, 2, ..., D$, we define edge utility mask matrix $\mathbf{M}_d \in \mathbb{R}^{|\mathcal{V} |\times |\mathcal{V} |}$ such that for any edge $(u,v) \in \mathcal{E}$, $\mathbf{M}_d (u,v) = \frac{1}{D}\mathbf{H}(d,u)\cdot \mathbf{H}(d,v) $. We also define the partition matrix $\mathbf{P} \in \mathbb{R}^{D\times |\mathcal{V}|}$ with elements $\mathbf{P}(d,v)= \sum_u \mathbf{M}_d(u,v)$. The \textit{orthogonality loss}, $\mathcal{L}_{orth}$, is designed to minimize the overlap of encoded structural information in different dimensions and is computed as follows:
\begin{align}
    \mathcal{L}_{orth} &= \text{MSE}\left(\frac{\mathbf{PP}^\text{T}}{\|\mathbf{PP}^\text{T}\|_F}, \frac{\mathbf{I}_D}{\|\mathbf{I}_D\|_F}\right),
\end{align}
where $\mathbf{I}_D$ is the identity matrix, and $\|\cdot\|_F$ is the Frobenius norm. 

The \textit{size loss} is introduced to prevent degenerate solutions that might arise from using only the orthogonality loss. For example, without this constraint, the model might assign all important edges to just one embedding dimension, making the others meaningless. To address this, we define a size variable for each dimension $d$ as $s_d=\sum_{u,v}\mathbf{M}_d(u,v)$. This represents the total importance (or summed weight) of all edges in the edge utility mask for dimension $d$. To encourage a balanced use of all dimensions, we maximize the entropy of the size variables $s_d$:
\begin{align}
    \mathcal{L}_{size} &= \log |D| + \sum_d \frac{s_d}{\sum_q s_q} \log \frac{s_d}{\sum_q s_q},
\end{align}
We added a small amount of noise to the input embedding matrix using isotropic Gaussian noise with $\sigma=0.2$. This nudges the auto-encoder to denoise embeddings during the transformation and also augments the training data set since the noise is not constant across epochs. We trained the auto-encoder with a learning rate of 0.1 and a batch size of 10,000 for 50 epochs.

In total, we aimed for 3 (walk length) $\times$ 2 (number of walks) $\times$ 3 (window size) $\times$ 3 (embedding dimension) $\times$ 3 (network year) = 162 DeepWalk representations from which 11 could not be created due to computational issues (6.79\%; training DeepWalk embeddings with 128 dimensions, 100 walks per node, and window size 8 exceeded our computational capacities in some cases). Together with 3 (embedding dimension) $\times$ 3 (network year) = 9 LINE representations, this led to 162 - 11 + 9 = 160 different embedding configurations.

\subsection{Prediction Performance}

We trained models to predict the outcome right-wing populist voting in the LISS panel sample using three feature sets. The first only contained the node embeddings of the LISS participants. The second feature set contained only individual covariates (see section Data), and the third feature set contained both embeddings and covariates.

For prediction, we used three different algorithms: Logistic regression with L2 regularization, k-nearest neighbor classification, and XGBoost classification. In total, we trained 2 (transformed yes/no) $\times$ 3 (embedding feature sets) $\times$ 160 (embeddings) $\times$ 3 (prediction algorithm) = 2880 predictive models for right-wing populist voting.

For logistic regression and k-nearest neighbor classification, missing continuous predictor values were imputed using the median. For all algorithms, missing categorical predictor values were added as a separate category.

We assessed the performance of the predictive models using nested stratified cross-validation (CV) with 5 inner and 5 outer folds. As the performance metric, we used the macro area under the curve (AUC) score for right-wing populist voting (unweighted average of AUC scores for each outcome class). We computed outcome class probabilities using one-vs-rest classification. We optimized hyperparameters of the predictive models using the average score in the inner CV loop using the Optuna framework \cite{akiba_optuna_2019} with 100 iterations. The hyperparameter ranges are specified in Supplementary Table S3.

To compare the performance of different predictive models, we estimated a Bayesian regression model to predict their outer CV loop macro AUC scores:
\begin{align*}
\text{score} &\sim \text{algorithm} \times \text{feature\_set} + \text{transformed} + \text{year} + \text{walk\_length} \times \text{num\_walks} \\ & + \text{window\_size} + \text{embedding\_dim}.
\end{align*}
Because AUC scores fall between 0 and 1, we used a generalized linear regression model with a beta family and logit link function. The nuisance parameter $\phi$ of the beta distribution was estimated as $\phi \sim \text{algorithm} \times \text{feature\_set}$ to allow for different variances in AUC scores between prediction algorithms and feature sets. Missing scores were estimated as additional parameters by the regression model. All independent variables were categorical. We used $\text{N}(0, 2.5)$ as weakly informative prior distributions for the regression coefficients and $\text{Exp}(1)$ on the parameters for estimating $\phi$. For non-applicable predictor combinations, we restricted regression coefficients to zero (e.g., only prediction models with embeddings in their feature set had embedding hyperparameter values). We assessed the posterior predictive distribution of the regression model and concluded that the predicted scores aligned sufficiently with the observed CV scores (see Supplementary Materials).

\subsection{Exploratory Analysis}

To investigate which connectivity patterns in the network are relevant for predicting our outcome variables, we used a modified version of the workflow described in \cite{piaggesi_dine_2024}: First, we retrained the prediction model with DINE-transformed embeddings and the best average prediction score for populist voting in the outer CV loop on the entire LISS panel data using only one stratified CV loop to optimize the hyperparameters as we did with previous models. Then, we calculated SHAP \cite{strumbelj_explaining_2014,lipovetsky_analysis_2001} feature importance scores for this model to assess how much each predictor contributed to the model predictions. We calculated SHAP values for each class and restricted the analysis to the scores for the populist class. From the predictors for each outcome, we selected the DINE-transformed embedding dimension with the highest mean absolute SHAP values for the subsequent analyses.

We calculated edge utility scores for the selected dimension and all edges in the network \cite{piaggesi_dine_2024}. The edge utility $\mu_d(u,v)$ quantifies the marginal contribution of dimension $d$ to the average element-wise product between the embeddings of the source and target node in the edge $(u, v)$ (see also Fig.~\ref{fig:workflow}C): 
\begin{align}
    \mu_d(u,v) = \Delta_D(u,v) -  \Delta_{D \symbol{92} \{d\}}(u,v),
\end{align}
where $\Delta_D(u,v)$ is the average embedding score of the edge, the average element-wise product between embeddings of nodes $u$ and $v$:
\begin{align}
    \Delta_D(u,v) = \frac{1}{D} \mathbf{H}(u) \cdot \mathbf{H}(v),
\end{align}
and $\Delta_{D \symbol{92} \{d\}}(u,v)$ the average embedding score excluding the dimension $d$.

Edge utility is positive (negative) when the similarity in dimension $d$ is above (below) the average similarity over all dimensions. Thus, edge utility is positive when either (a) all dimensions are similar in sign and strength but $d$ has a higher value than the other dimensions or (b) dimension $d$ is similar in sign and strength but the other dimensions are dissimilar. In contrast, edge utility is negative when either (a) all dimensions are similar in sign and strength but $d$ has a lower value than the other dimensions or (b) dimension $d$ is dissimilar in sign and strength but the other dimensions are similar. We interpret edge utility as a measure of how important edges are for predicting right-wing populist voting, given that it is highly associated with the selected embedding dimension.

We aggregated edge utility at the person-level by defining edge utility strength as the sum of first-order edge utilities in dimension $d$:
\begin{align}
    S_d(u) = \sum_{v}{\mu_d(u,v)}.
\end{align}
Values different from zero indicate that a person has first-order relations that are important for predicting right-wing populist voting. We interpret edge utility strength as an indicator of differences in local network structure that are predictive of right-wing populist voting.

Lastly, we aggregated edge utility at the municipality level (in total 342 municipalities) by averaging over edges that have source and target nodes in the same or a different municipality:
\begin{align}
    \gamma_d(m,n) = \frac{1}{|\mathcal{E}'|}\sum{\mu_d(u_m,v_n)},
\end{align}
where $\mathcal{E}'$ is the subset of edges that connect persons from municipalities $m$ and $n$.
We then calculated the average edge utility strength for each municipality:
\begin{align}
    Q_d(m) = \sum_n{\gamma_d(m,n)}.
\end{align}
We correlated municipality-level average edge utility strength with aggregated municipality statistics: Mean income per capita, median household assets, address density, population share in three age groups (0-15, 14-25, $>$ 64 years), population share of highest achieved educational degree (lower, intermediate, higher), and population share with a non-western migration background. Additionally, we obtained the percentage of votes for populist parties in each municipality in the Netherlands for the parliamentary election on November 22nd, 2023.

\paragraph{Acknowledgements}
We would like to thank ODISSEI (the Open Data Infrastructure for Social Sciences and Economic Innovations) for making administrative registry data more accessible. ODISSEI also provided data access for this project (grant number 184.035.014).
The LISS panel data were collected by the non-profit research institute Centerdata (Tilburg University, the
Netherlands). Funding for the panel's ongoing operations comes from the Domain Plan SSH and ODISSEI
since 2019. The initial set-up of the LISS panel in 2007 was funded through the MESS project by the
Netherlands Organization for Scientific Research (NWO). Javier Garcia-Bernardo acknowledges support from 
NWO (grant number VI.Veni.231S.148).

\section*{Funding}
Funding was provided by ODISSEI through a microdata access grant awarded to MK (grant number 184.035.014) and by the Netherlands Organization for Scientific Research (NWO) to JGB (grant number VI.Veni.231S.148).

\section*{Competing Interests}
The authors have no competing interests to declare.

\section*{Ethics Statement}

Ethics approval was obtained from the Human Research Ethics Committee at Delft University of Technology on June 9th 2024 (application number 4403). All analyses on raw registry data were conducted in the secure remote access environment by Statistics Netherlands (project number 9780). In this environment, all personal data is pseudonomized. All results presented in this study were checked by Statistics Netherlands for risks of personal information disclosure and approved for publication (for details on privacy, see \url{https://www.cbs.nl/nl-nl/over-ons/dit-zijn-wij/onze-organisatie/privacy}). To prevent disclosure of personal information, only aggregated, fully anonymized results are presented in this study.

\section*{Consent for Publication}

Not applicable.

\section*{Data Availability}

The raw datasets used in this study are not openly available. The Dutch population network and registry data can be accessed through Statistics Netherlands under certain conditions (for details, see \url{https://www.cbs.nl/nl-nl/onze-diensten/maatwerk-en-microdata/microdata-zelf-onderzoek-doen}). The LISS Panel data can be accessed through the LISS Archive (see \url{https://www.lissdata.nl/how-it-works-archive}). The embeddings that were created as part of this study are archived in the Data Storage Facility from Statistics Netherlands and ODISSEI. They can also be accessed through Statistics Netherlands with consent from the authors. The metadata for the embeddings can be found here:

\begin{itemize}
    \item[-] DeepWalk embeddings: \url{https://doi.org/10.34894/VRNLKJ}
    \item[-] DeepWalk embeddings DINE transformed: \url{https://doi.org/10.34894/DRRGIV}
    \item[-] LINE embeddings: \url{https://doi.org/10.34894/RVVTRB}
    \item[-] LINE embeddings DINE transformed: \url{https://doi.org/10.34894/9JLZ4E}
\end{itemize}

Aggregated data that has been checked for risks of personal information disclosure and approved by Statistics Netherlands is available at \url{https://github.com/odissei-explainable-network/netaudit}.

Municpality statistics are openly available via CBS Open Data StatLine (\url{https://opendata.cbs.nl/statline/portal.html?_la=nl&_catalog=CBS&tableId=85039NED&_theme=246}). Municiaplity-aggregated voting data are openly available via the Overheid.nl website  (\url{https://data.overheid.nl/dataset/verkiezingsuitslag-tweede-kamer-2023}).

\section*{Materials Availability}

Not applicable.

\section*{Code Availability}

Code is available at \url{https://github.com/odissei-explainable-network/netaudit}.

\section*{Author Contributions}

ML, JGB, and MK conceptualized the study. MK acquired funding and supervised the project together with JGB. ML, SD, and FH wrote the software necessary for executing the study. ML conducted the data analysis. ML, JGB, FH, and MK wrote the manuscript which was reviewed and approved by all authors.

\bibliographystyle{naturemag}

\end{document}